\documentclass[aps,pra,twocolumn]{revtex4-1}

\usepackage{amsmath}
\usepackage{amsthm}
\usepackage{graphicx}
\usepackage{xcolor}
\usepackage{amsfonts}
\usepackage{cleveref}
\usepackage{ulem} 

\begin{document}

\title{Squeezing of intensity noise in nanolasers and nanoLEDs}
\author{Jesper Mork and Kresten Yvind} 
\affiliation{Center for Nanophotonics, DTU Fotonik, Technical University of Denmark, DK-2800 Kgs. Lyngby, Denmark}
\email{jesm@fotonik.dtu.dk}

\date{\today}	

\begin{abstract}
The intensity noise of nanolasers and nanoLEDs is analyzed for conventional and quiet pumping of electrons into the active region. It is shown that nanoLEDs with Purcell-enhanced emission rates, enabled by extreme dielectric confinement, may be excellent sources of intensity-squeezed light within a bandwidth of several gigahertz. The physics of intensity squeezing is elucidated and it is shown that the parameter dependence can be explained by an effective Fano factor.
\end{abstract}

\maketitle

In many applications of lasers, it is important to reduce the noise of the laser signal. The laser usually displays Poisson intensity statistics, corresponding to the laser being well approximated by a coherent state. This means that the signal-to-noise ratio is entirely determined by the power level, and leads to a minimum required power level, e.g. for optical transmission systems, to avoid bit-errors \cite{Saleh1992}. Intensity-squeezed light, on the other hand, allows to reach the maximum capacity of a link \cite{Saleh1992}. It was shown by Yamamoto et al.  \cite{Machida1987,Yamamoto1987} that a semiconductor laser can generate sub-poissonian intensity-squeezed light by driving the laser in a constant-current-mode.  The degree of squeezing achieved in semiconductor lasers so-far \cite{Machida1988,Richardson1991,Kilper1996,Kaiser2001} is, however, severely limited by a number of factors, including weak side modes \cite{VanDerLee2000a}, intrinsic losses in the laser cavity \cite{Tromborg1994,Gallion1997}, and current leakage \cite{Maurin2005}. Furthermore, intensity-squeezed output is observed within a small bandwidth, which is typically insufficient for transmission of data  

In this paper we study the fundamental noise properties of nanolasers and nanoLEDs and show that nanoLEDs can be operated in a regime where the intensity noise is strongly squeezed, overcoming intrinsic limitations of conventional macroscopic devices. Recent years have witnessed tremendous progress in nanofabrication technology, allowing the realization of a new class of microlasers and nanolasers \cite{Matsuo2010,Jang2015,Ota2017a,Kreinberg2017,Yu2017,Crosnier2017a} with spontaneous emission factor, $\beta$, approaching unity \cite{Lermer2013,Ota2017a}. A $\beta$-value of 1 implies that all spontaneous emission noise is channelled into the cavity mode, and it might be expected that such devices  would show increased noise. Instead we show here that a near-unity value of $\beta$ is key to squeezing the intensity noise in nanoLEDs. Furthermore, by exploiting new cavity designs for extreme dielectric confinement \cite{Hu2016,Choi2017,Wang2018,Hu2018c}, we show that squeezing can be obtained over a bandwidth of several gigahertz even at very low power levels. 

These results provide new insight into the noise properties of devices with near-unity $\beta$-factor, and break with the mantra that high power is required for realizing low-noise signals. NanoLEDs could thus become key devices for realizing on-chip optical interconnects with low energy-consumption \cite{Miller2017a}. In contrast to recent theoretical work addressing the quantum noise of microlasers and nanolasers \cite{Auffeves2011,Strauf2011f,Moelbjerg2013,Lorke2013,Gies2017,Kreinberg2017,Mork2018}, we here take into account the quantum statistics of the cavity-outcoupling process, which is essential in order to describe intensity noise squeezing. In terms of methodology, we extend a recently suggested stochastic approach \cite{Puccioni2015,Mork2018} to include the process of "quiet" pumping. The results are compared to analytical results derived from Langevin equations. 

We model the dynamics of nanolasers and nanoLEDs by the following set of rate equations \cite{Mork2018} for the number of excited emitters, $n_e(t)$, and the number of photons in the cavity, $n_p(t)$,
\begin{align}
\frac{dn_e}{dt} &= R_p(t)-\gamma_r (2n_e-n_0)n_p-\gamma_tn_e, \label{eq:dnedt}\\
\frac{dn_p}{dt} &= \gamma_r (2n_e-n_0)n_p+\gamma_r n_e -\gamma_cn_p \label{eq:dnpdt}.
\end{align}
Here, $n_0$ is the total number of emitters (quantum dots), $\gamma_r$ is the coupling rate between a single-emitter and a photon in the cavity mode, which can be expressed in terms of fundamental material and cavity parameters \cite{Mork2018}, $\gamma_t$ is the total emitter decay rate, and $\gamma_c=\gamma_{\rm out}+\gamma_{\rm int}$ is the cavity decay rate, with $\gamma_{\rm out}$ being the coupling rate into the output channel and $\gamma_{\rm int}$ being all other cavity losses, e.g. due to residual absorption and disorder.

The term $R_p(t)$ is the rate at which emitters are excited by the injected current. Its description depends on how the laser is pumped. For conventional pumping, realized by a current source with classical Poisson-distributed shot-noise, we take $R_p(t)=R_{p,\text{conv}}(t)=\gamma_p(n_0-n_e(t))$. This expression is appropriate for a finite number of emitters, where state-filling dictates $n_e \le n_0$ and the injection efficiency decreases as more emitters are excited \cite{Mork2018}. The total external pump rate in this case is given by $\gamma_p n_0$, with the rate $\gamma_p n_e$ representing leakage. A more accurate model could take into account an upper pump reservoir, e.g. constituted by carriers in the wetting layer of a quantum dot laser, but the present model already elucidates important physics related to the squeezing process. For the case of quiet pumping, $R_p(t)=R_{p,\text{quiet}}(t)$, the pump is a periodic injection of electrons at constant rate, corresponding to sub-poissonian electron statistics and unity injection efficiency. 

Conventionally, quantum noise in semiconductor lasers is modelled by the addition of Langevin white-noise forces, $F_n(t)$ and $F_p(t)$, to the RHS of Eqs. (\ref{eq:dnedt})-(\ref{eq:dnpdt}), with the magnitude of the noise being described through diffusion coefficients, $2D_{xy}=\langle F_x(t)F_y(t) \rangle$ ($x,y=n_e,n_p$), see e.g. Ref. \onlinecite{Coldren2012}. An alternative stochastic approach has been introduced \cite{Puccioni2015,Mork2018}, where the noise instead enters by demanding that the number of excited emitters, $n_e(t)$, and the photon number, $n_p(t)$, be integer-valued. This implies that the various terms in Eqs. (\ref{eq:dnedt})-(\ref{eq:dnpdt}) are to be considered as stochastic processes, which can be taken to be Poisson-distributed during small time-steps of length $\Delta t$, into which the total simulation time is divided. It was recently shown that the intracacity photon number noise obtained by this approach agrees well with analytical results obtained from the Langevin approach \cite{Mork2018}, at least when the number of emitters exceeds $\sim 10$. 
 
Here, we explicitly consider the out-coupling process as an additional random process.  In the stochastic simulations, conventional shot-noise-limited pumping is simulated as all other reservoir-exchanges \cite{Mork2018}, whereas for "quiet" pumping, electrons are injected with period  $1/R_{p,\text{quiet}}$, requiring that
$(R_{p,\text{quiet}}\Delta t)^{-1}$ is an integer $M \ge 1$. We typically use $M \sim 100$ and check that the noise measures have converged.

\begin{figure}[ht!]
	\includegraphics[width=0.23\textwidth]{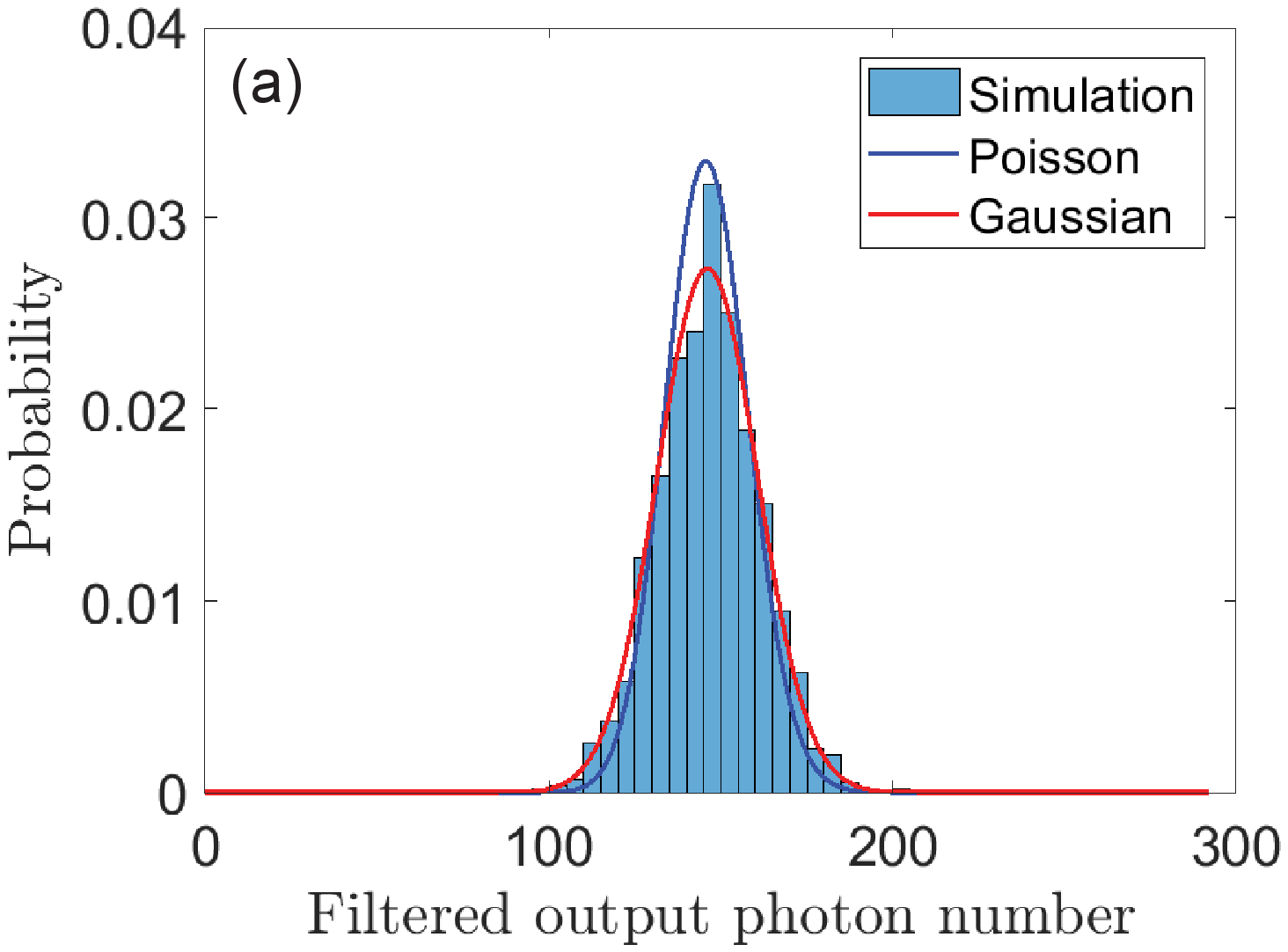}
 	\includegraphics[width=0.23\textwidth]{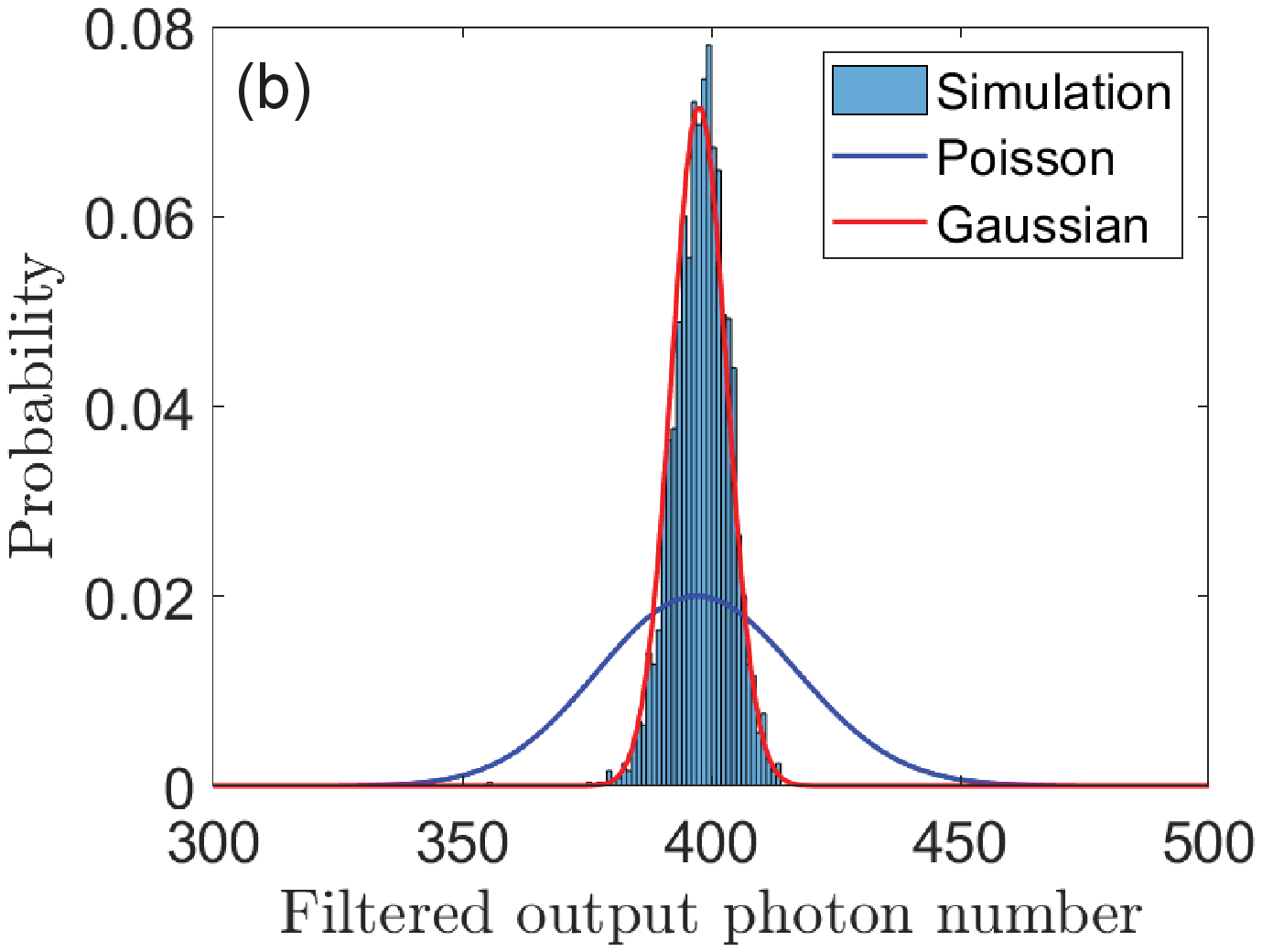}
	\caption{Probability distribution functions of output photon number in  400 ps time slots for a nanolaser with (a) conventional pumping and (b) quiet pumping. The red curves are gaussian fits and the blue curves are Poisson distributions with the same mean value as the simulations.}
	\label{fig:Histograms-laser}
\end{figure}

We first consider the case of a nanolaser, using parameter values representative of the quantum dot photonic crystal laser investigated in Ref. \onlinecite{Ota2017a}: $n_0=50,\ \beta=0.97,\ \gamma_t=5\times 10^{9} {\rm s}^{-1}$, and $\gamma_{cav}=0.75\times 10^{11} {\rm s}^{-1}$ (corresponding to a cavity Q-factor of 16215 at the considered wavelength of $\lambda=1.55\ \mu$m).
The histograms in Figs. \ref{fig:Histograms-laser} show detected photon number distributions for at $R_p=1\times 10^{12} {\rm s}^{-1}$ for (a) conventional and (b) quiet pumping. If the laser is used for generating low-noise pulses ("bits") at a bit rate of $B$, the relevant noise to consider corresponds to having a photodetector with an integration time of $T_p=1/B$. In Fig. \ref{fig:Histograms-laser} we consider a bit rate of 2.5 Gb/s and model the detection by a temporal (square) filter of duration 400 ps. 

For the case of noisy pumping, Fig. \ref{fig:Histograms-laser}(a), the simulated distribution is well fitted by a gaussian probability distribution, with a FWHM close to that of the corresponding Poisson distribution with the same mean.
On the other hand, quiet pumping, Fig. \ref{fig:Histograms-laser}(b), results in a photon number distribution which is clearly sub-poissonian. The lower average photon number obtained for conventional pumping is due to reduced pump-efficiency, which can be compensated for by using a larger pump, but has  no effect on the conclusions. 

\begin{figure}[ht!]
\includegraphics[width=0.23\textwidth]{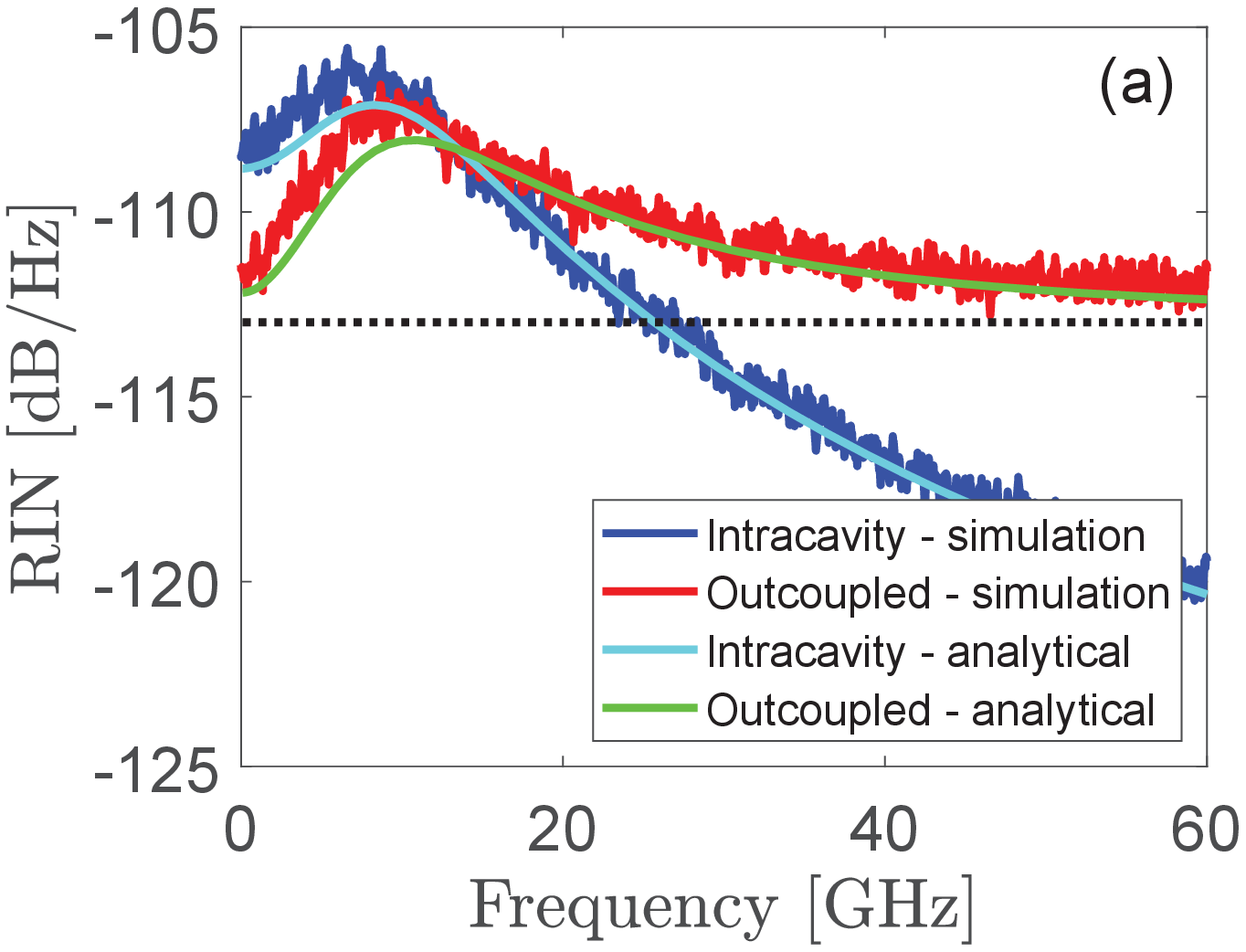}
\includegraphics[width=0.23\textwidth]{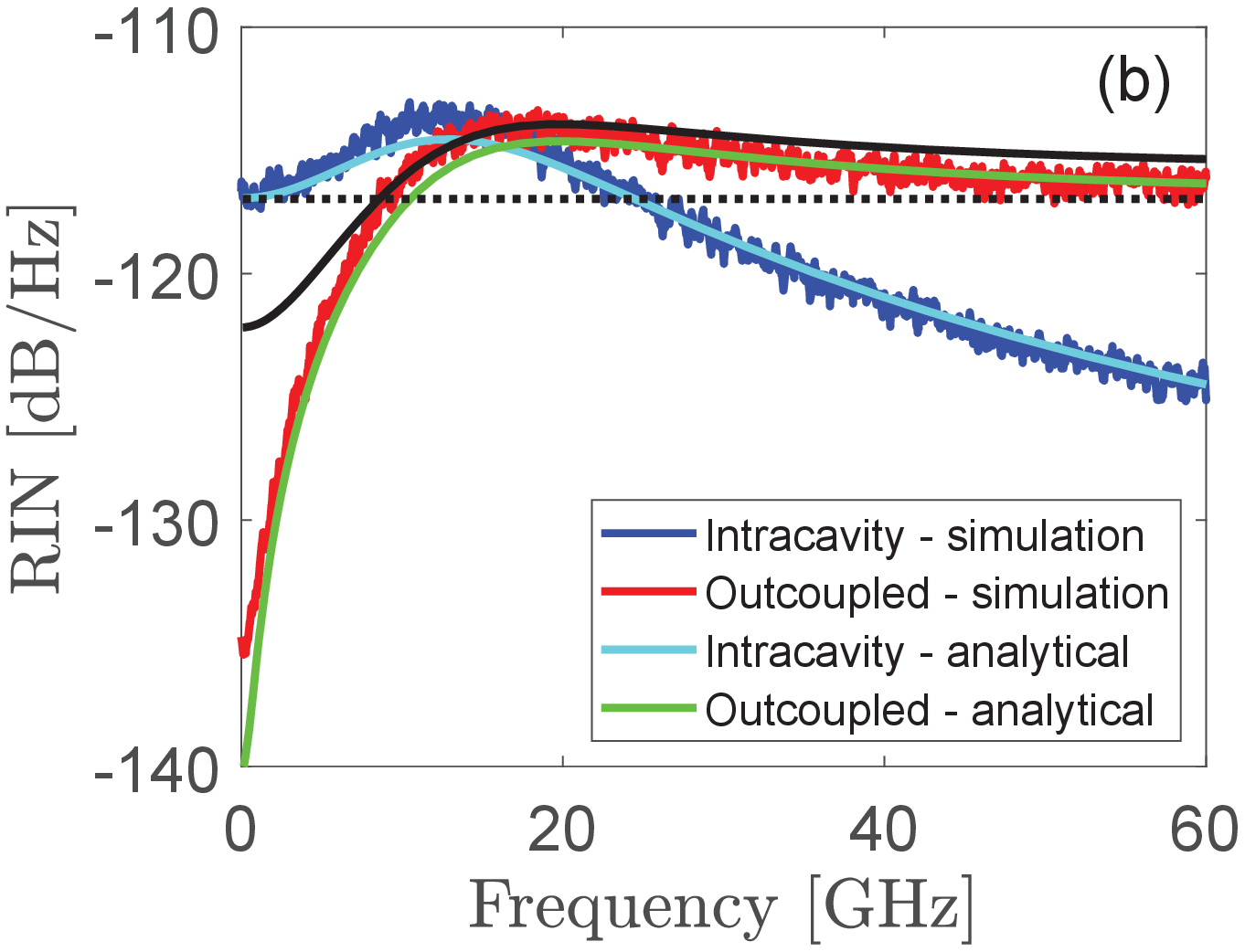}
\caption{RIN spectra of nanolaser for (a) conventional and (b) quiet pumping, for outcoupled (red) and intracavity (blue) photons. The parameters are the same as in Fig. \ref{fig:Histograms-laser}. The black dashed line shows the standard quantum limit of $2\hbar\omega_0/\bar{P}_{\rm out}$ and the black solid line is the analytical result for a finite internal loss of $2 {\rm cm}^{-1}$}.
\label{fig:RIN-spectra-laser}
\end{figure}

In order to get insight into the mechanism of squeezing, Figs. \ref{fig:RIN-spectra-laser} show the relative intensity noisy (RIN) spectra for noisy and quiet pumping, comparing external, measurable, spectra to the intracavity spectra. The analytical predictions are obtained by a small-signal analysis of Eqs.  (\ref{eq:dnedt})-(\ref{eq:dnpdt}), including Langevin noise terms:   
\begin{align}
\text{RIN}_{\text{out}}(\Omega)&= \text{RIN}_{\text{int}}(\Omega)+ \nonumber \\
& \frac {2\hbar\omega_0}{\bar{P}_{out}}
\left[1-2\eta_0\gamma_c\frac{\Gamma_{ee}(\omega_R^2-\Omega^2)+\Gamma\Omega^2}{|D(\Omega)|^2}\right],
\label{eq:RINout}
\end{align}
with the intracavity noise given by
\begin{equation}
\text{RIN}_{\text{int}}(\Omega)=4\frac{(\Gamma_{ee}^2+\Omega^2)D_{pp}+\Gamma_{pe}^2 D_{ee}+2\Gamma_{pe}\Gamma_{ee} D_{pe}}{\bar{n}_p^2|D(\Omega)|^2}.
\label{eq:RINin}
\end{equation}
Here, $\bar{n}_p$ and $\bar{n}_e$ are the steady-state values of photon number and emitter excitation, the differential damping rates are given as 
$\Gamma_{ee} = \epsilon\gamma_p+\gamma_t+2\gamma_r \bar{n}_p$, $\Gamma_{ep}=\gamma_r(2\bar{n}_e-n_0)$,
$\Gamma_{pe} =\gamma_r(2\bar{n}_p+1)$, $\Gamma_{pp}=\gamma_c-\gamma_r(2\bar{n}_e-n_0)$, $\Gamma=\Gamma_{ee}+\Gamma_{pp}$, and the diffusion coefficients as 
$2D_{ee} = \epsilon \gamma_p(n_0-\bar{n}_e)+\gamma_r n_0 \bar{n}_p+\gamma_t\bar{n}_e$, 
$2D_{pp} = \gamma_r n_0\bar{n}_p+\gamma_r\bar{n}_e+\gamma_c\bar{n}_p$, and $2D_{ep}=2D_{pe}=-\gamma_r n_0\bar{n}_p-\gamma_r\bar{n}_e$. Furthermore, the system determinant is 
$D(\Omega)=-\Omega^2-i\Gamma\Omega+\omega_R^2$, with intrinsic relaxation oscillation frequency 
$\omega_R^2 = \Gamma_{ep}\Gamma_{pe}+\Gamma_{ee}\Gamma_{pp}$, and the outcoupled power is $\bar{P}_{\rm out}=\hbar\omega_0\gamma_{\rm out}\bar{n}_p$. The extraction efficiency, i.e., the ratio of out-coupling rate and total cavity loss rate is $\eta_c=(\gamma_c-v_g\alpha_{\rm int})/\gamma_c$, with $\alpha_{\rm int}$ being the internal loss, and $v_g$ the group velocity. In these expressions, $\epsilon=1\ (0)$ for standard shot-noise-limited (quiet) pumping. Compared to previous theoretical analyses \cite{Yamamoto1987,Tromborg1994,Gallion1997,VanDerLee2000a}, we have accounted for near-unity spontaneous emission factor $\beta=\gamma_r/\gamma_t$, and a finite number of emitters, as appropriate when analyzing nanolasers.

Simulated and analytical RIN-spectra in general agree very well. 
In the case of shot-noise-limited pumping, Fig. \ref{fig:RIN-spectra-laser}(a), the RIN of the out-coupled laser signal is reduced by 3 dB compared to the intra-cavity laser power at low frequencies and approaches the standard quantum limit of $2\hbar\omega_0/\bar{P}_{\rm out}$ at frequencies exceeding the characteristic rates describing the laser dynamics. The internal RIN continues to decrease with frequency due to the filtering imposed by the laser dynamics, as expressed by the response function  $|H(\Omega)|^2=|D(\Omega)/D(0)|^2$.  

Figs. \ref{fig:RIN-spectra-laser} clearly show that the out-coupling process reduces the intensity noise at low frequencies, both for conventional and quiet pumping. 
This may appear surprising considering the random nature of the outcoupling process: Individual photons are either transmitted through the laser output mirror or reflected back into the cavity. Yamamoto et al. explain the noise reduction as being due to destructive interference with external vacuum fluctuations being reflected off the laser mirror \cite{Yamamoto1987}. The noise reduction may, however, also be seen as the result of the anti-correlation between photons inside the cavity and photons coupled out of the cavity: If a photon is coupled out, the number of internal photons decreases and the probability of coupling out yet another photon decreases. This anti-correlation, expressed by the negative sign in the last term of (\ref{eq:RINout}), imparts an anti-bunching mechanism to the stream of output photons and the intensity correlation, $g^{(2)}(0)$, as well as the intensity noise decreases. Alternatively, the squeezing may be understood as the result of the lossless transfer of the sub-poissonian stream of input electrons to a sub-poissonian stream of output photons \cite{Chusseau2003}. At low frequencies, corresponding to long observation times, it is guaranteed that a given number of electrons in the input electron-stream results in the same number of photons in the output. In contrast, the intracavity energy can be distributed among the photons and the emitters, and the fluctuations of each of these populations may be large.  

The black solid line in Fig. \ref{fig:RIN-spectra-laser} shows the effect of increasing the internal losses to $2 {\rm cm}^{-1}$ ($\gamma_{int}=1.71\times 10^{10} {\rm s}^{-1}$ and $Q=71000$), corresponding to $\eta_c=0.77$. Although this is a small loss, it is seen to strongly reduce the squeezing. Since the small gain region in nanolasers dictates a small out-coupling loss, in order to achieve lasing, internal losses make it difficult to realize nanolasers with strong noise suppression. Instead, we shall show that nanoLEDs allow to overcome this limit.

\begin{figure}[ht!]
\includegraphics[width=0.23\textwidth]{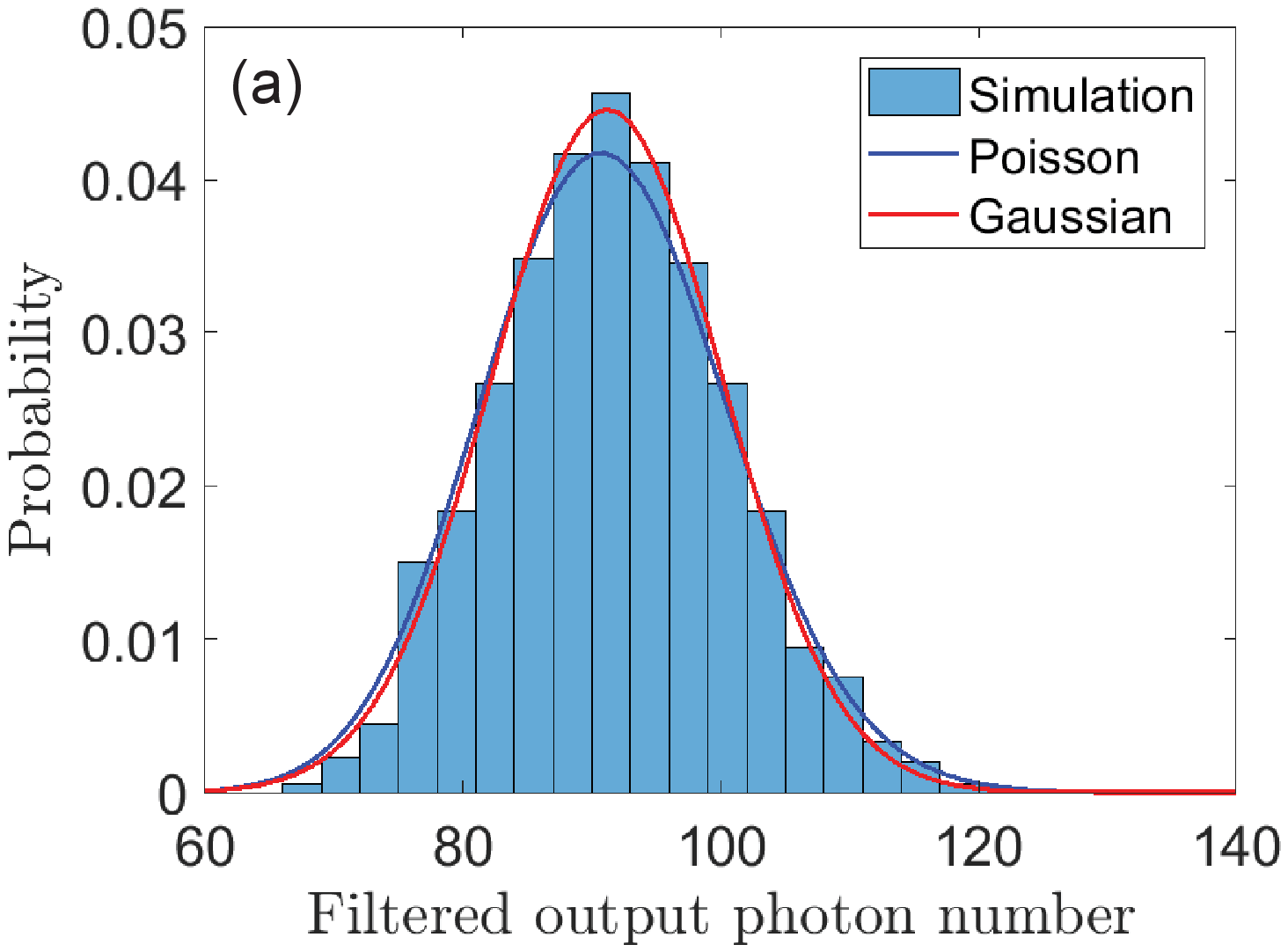}
\includegraphics[width=0.23\textwidth]{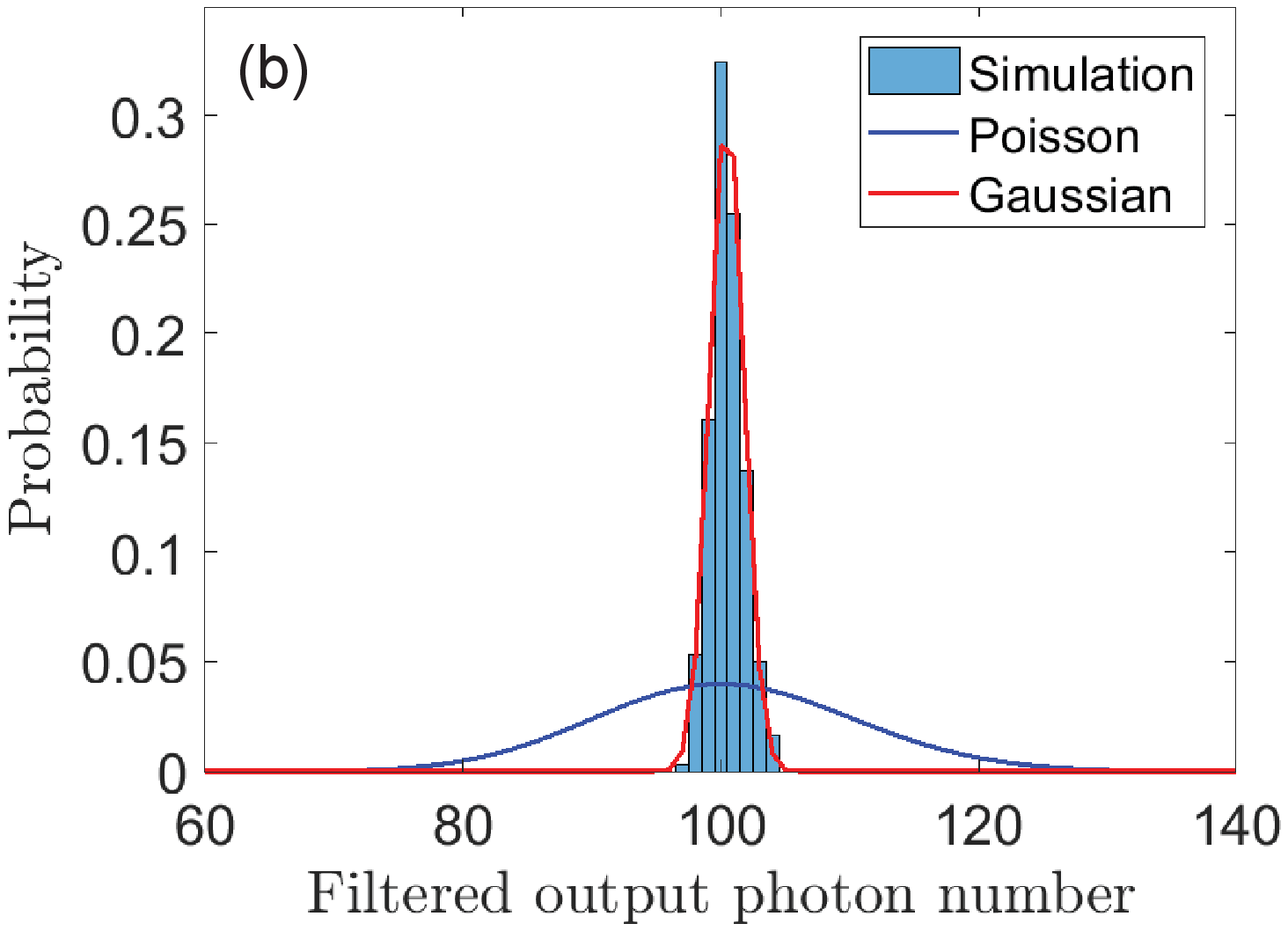}
\caption{Probability distribution functions (histograms) of the number of output photons in 100 ps timeslots for a nanoLED with (a) conventional and (b) quiet pumping. The red curves are gaussian fits and the blue curves are Poisson distributions.}
\label{fig:Histograms-LED}
\end{figure}
\begin{figure}[ht!]
\includegraphics[width=0.23\textwidth]{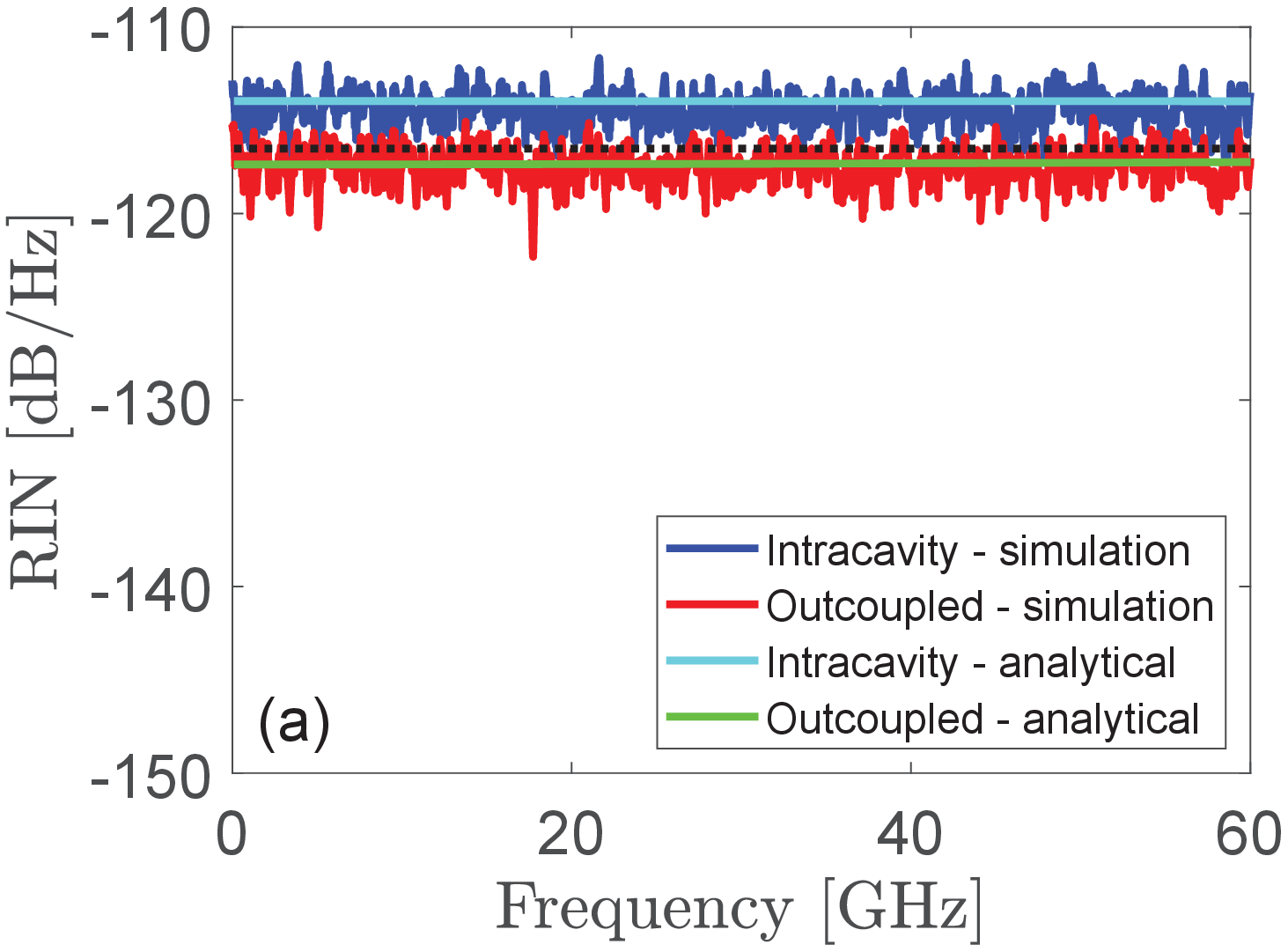}
\includegraphics[width=0.23\textwidth]{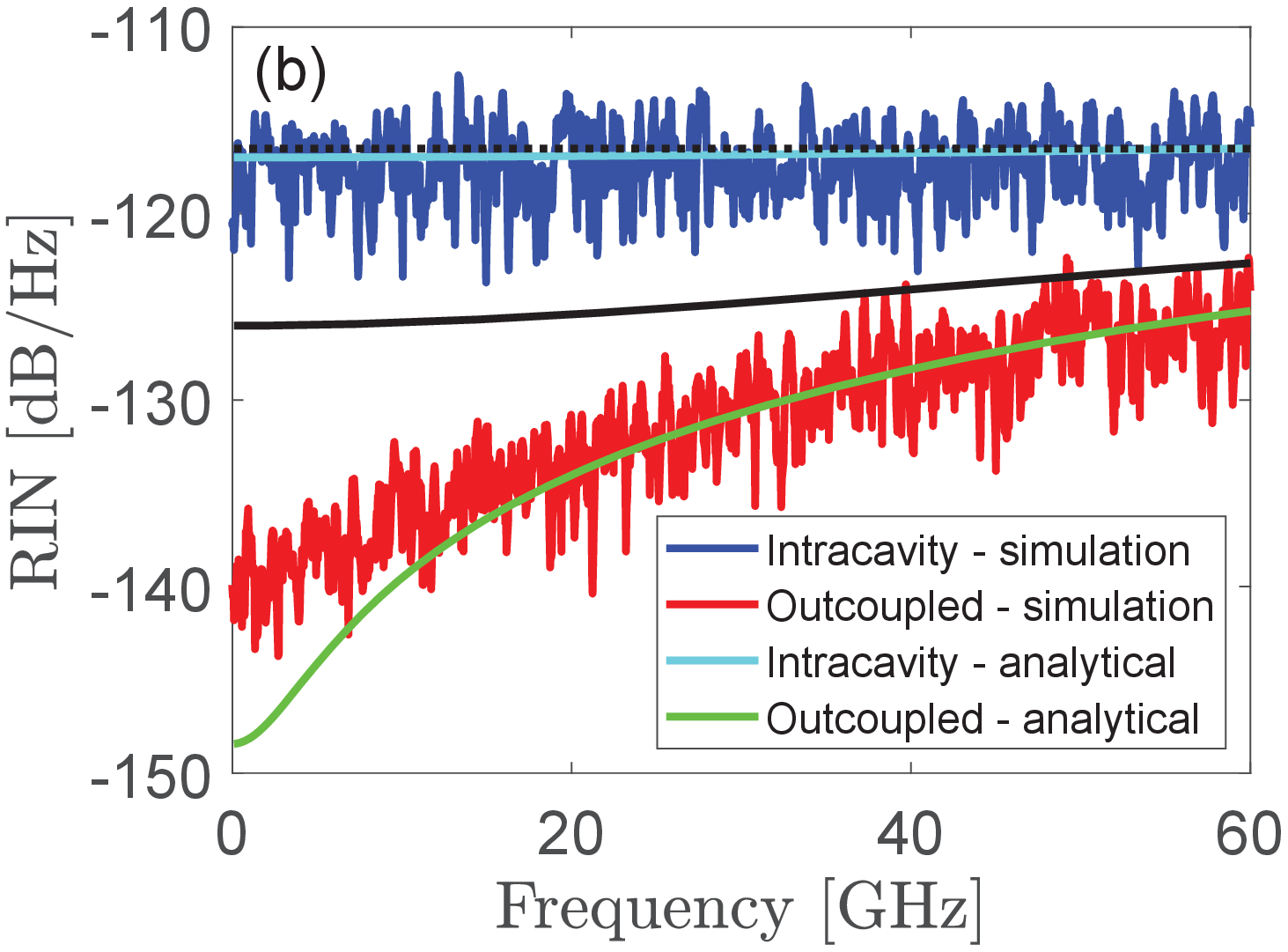}
\caption{RIN spectra of a nanoLED for (a) conventional and (b) quiet pumping, for outcoupled (red) and intracavity (blue) photons.  The black solid line shows the analytical RIN of the out-coupled power for a reduced  $\beta=0.9$, and the black dashed line shows the standard quantum limit.}
\label{fig:RIN-spectra-LED}
\end{figure}

Fig. \ref{fig:Histograms-LED} shows photon number distributions for a nanoLED as detected in bit slots of duration 100 ps, corresponding to operation at 10 Gb/s, and Fig. \ref{fig:RIN-spectra-LED} shows the corresponding RIN spectra. The parameters used in Fig.  \ref{fig:Histograms-LED} are: 
$N_0=10$, $\beta=1$, $\gamma_t=1\times 10^{12} {\rm s}^{-1}$, $\gamma_{cav}=6\times 10^{13} {\rm s}^{-1}$, and $\gamma_{int}=4.3\times 10^{10} {\rm s}^{-1}$. Compared to the nanolaser case, the emitter-cavity coupling rate, $\gamma_r$, has been considerably increased. Such enhanced coupling can be achieved through the Purcell effect, since the spontaneous emitter decay rate scales inversely with the cavity mode volume \cite{Purcell1946}. 
It was recently shown, that cavity mode volumes much smaller than the usual diffraction limit can be achieved in dielectrics by using new cavity designs \cite{Hu2016,Choi2017,Wang2018,Hu2018c}. Such extreme dielectric confinement should allow enhancing the emitter cavity-coupling rate by several orders of magnitude. 

The nanoLED RIN spectra show less frequency dependence than the nanolaser due to the large emitter-cavity coupling rate and large cavity decay rate. The latter also results in a small population of intracavity photons, and the transfer function has a an amplitude smaller than 1, meaning that the noise is not enhanced beyond the value given by the standard quantum limit. in contrast to the nanolaser, cf. Fig. \ref{fig:RIN-spectra-laser}. The reduced squeezing seen at low frequencies for the simulated results as compared to the analytical results is attributed to dynamical pump blocking and the resulting leakage.
Fig. \ref{fig:RIN-spectra-LED} also shows the RIN spectrum  for a reduced $\beta$-factor of 0.9, demonstrating that this is a critical parameter for nanoLEDs and needs to be close to unity to achieve strong squeezing. 

\begin{figure}[ht!]
	\includegraphics[width=0.23\textwidth]{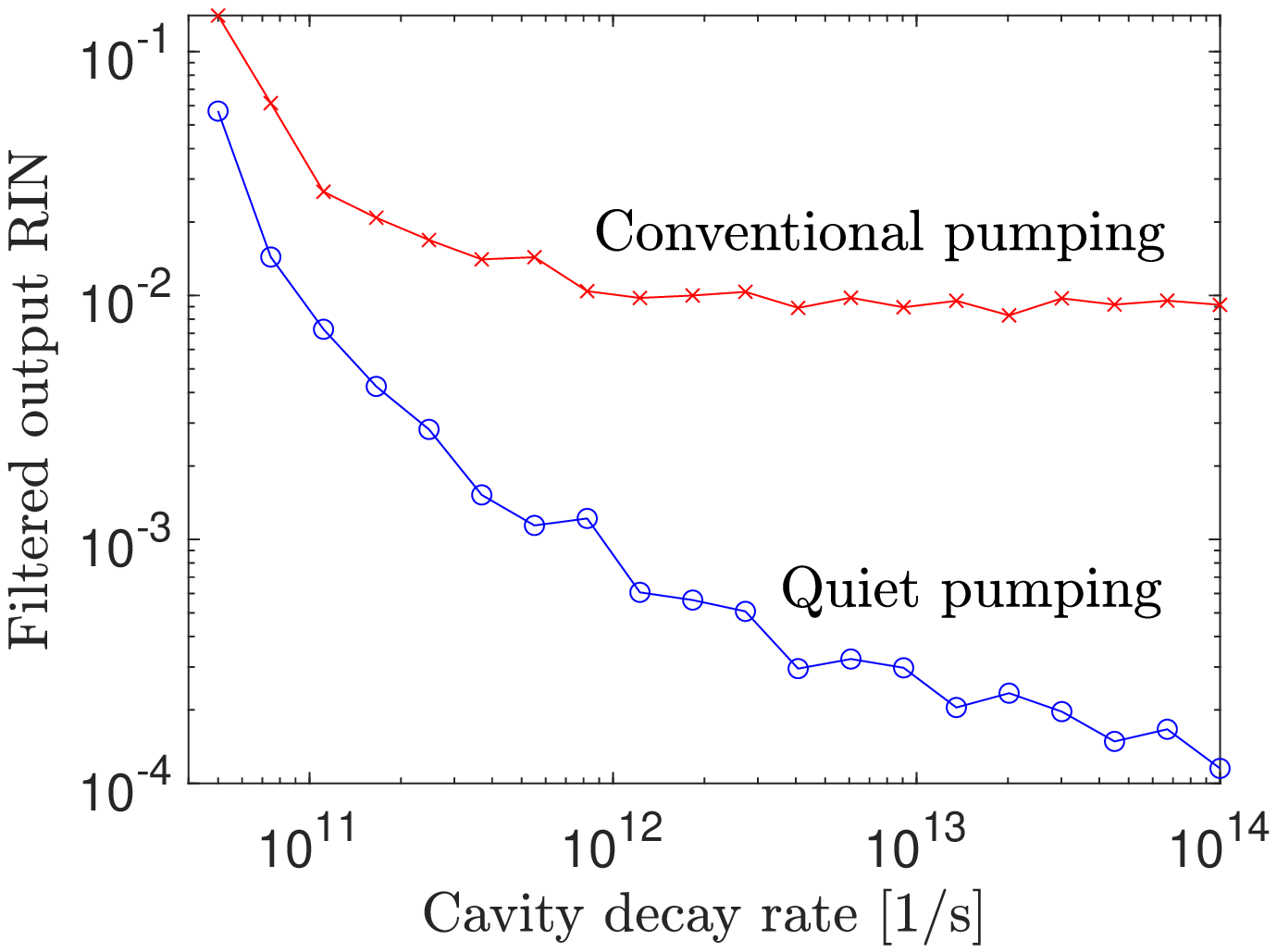}
	\includegraphics[width=0.23\textwidth]{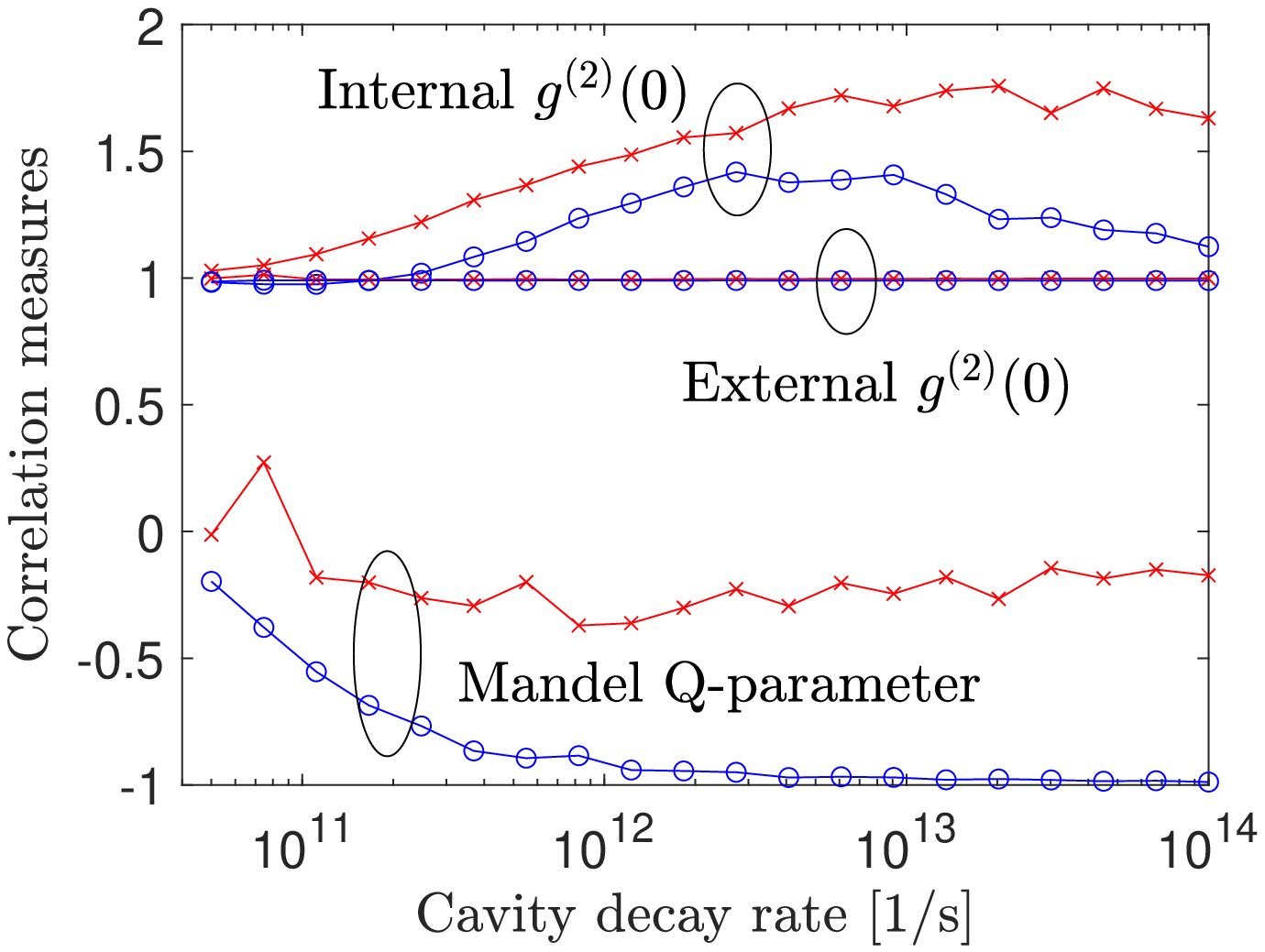}
	\caption{Different measures of the intensity noise in dependence of the cavity decay rate. Blue (red) curves are for quiet (conventional) pumping. The measurement bandwidth is 10 GHz. Other parameters are the same as in Fig. \ref{fig:RIN-spectra-LED}.}
	\label{fig:gam_cav_variation}
\end{figure}

In order to compare nanolasers and nanoLEDs, Fig. \ref{fig:gam_cav_variation} shows different measures of the  intensity quantum noise in dependence of the cavity decay rate, $\gamma_{\rm cav}$. As $\gamma_{\rm cav}$ increases, a transition occurs from lasing to LED-operation. This is evidenced by the zero-delay intensity correlation for the intracavity photon distribution, $g^{(2)}(0)$, which changes, upon increasing $\gamma_{\rm cav}$, from a value of 1, indicating coherent light, to a value of 2, indicating thermal statistics. In the case of quiet pumping and  very high $\gamma_{\rm cav}$, there appears to be a recovery of Poisson statistics for the intracavity photons. We attribute this to the absence of any re-absorption events, before the photons are coupled out. The variation of the RIN with $\gamma_{\rm cav}$ clearly shows the advantage of operating in the nanoLED regime. Whereas the RIN  quickly saturates for conventional pumping, it decreases with $\gamma_{\rm cav}$ for quiet pumping.  


The correlation of photons detected externally, with a detector response time of 100 ps, $g_{\rm ext}^{(2)}(0)$, hardly changes with cavity decay rate and is almost independent of whether conventional or quiet pumping is used. From the relation $\langle\Delta n_p^2\rangle=[ g_{\rm ext}^{(2)}(0)-1 ] \langle n_p \rangle^2+\langle n_p \rangle$, it is thus seen that if the average photon number is large (corresponding, e.g, to a long detector response time), even a small reduction of $g_{\rm ext}^{(2)}(0)$ below unity indicates strongly sub-poissonian intensity fluctuation.
On the other hand, Mandel's Q-factor \cite{Loudon2005}, $Q_{\rm M}=\langle n_p \rangle\left[ g_{\rm ext}^{(2)}(0)-1 \right]$ is seen to provide a sensitive measure of squeezing,  
with  $Q_{\rm Ml}=-1$ indicating the maximum degree of squeezing.



The squeezing obtained in nanolasers and nanoLEDs in dependence of device parameters can be summarized by an effective Fano factor. Consider a quiet input electron stream consisting of a sequence of time (bit) slots of duration $T_B$, each containing exactly $N$ electrons (number states). If these number states are transferred to the detector with efficiency $\eta$, the resulting probability distribution is a binomial counting distribution with mean $\langle n \rangle=\eta N$, variance $\sigma^2=\eta(1-\eta)N$, and Fano factor $F=\sigma^2/\langle n \rangle=1-\eta$ \cite{Saleh1992}. It is easily seen that in this case ${\rm RIN}_{\rm quiet}/{\rm RIN}_{\rm conv}=F$, so that the number states are conserved for $F=0$ ($\eta=1$). 

The efficiency $\eta$ has contributions of different origin: $\eta=\eta_{\rm c}\eta_{\rm dyn}\eta_{\rm det}$, with $\eta_{\rm c}$ being the efficiency by which electrons are converted into out-coupled photons,  $\eta_{\rm dyn}$ the dynamical transfer efficiency, depending on the bit duration $T_B$, and $\eta_{\rm det}$ the detector efficiency including link losses (assumed to be unity in the simulations). For a laser operating well-above threshold, $\eta_{\rm c}=\eta_{\rm i}(\gamma_{\rm c}-\gamma_{\rm int})/\gamma_{\rm c}$, while for an LED, $\eta_{\rm c}=\eta_{\rm i}\beta$. Here, $\eta_{\rm i}$ is the pump efficiency, and $\eta_{\rm dyn}\simeq |H(\xi/T_b)|^2$, where $\xi$ is a number of order unity depending on the temporal waveform and the variation of the RIN within the signal bandwidth. This implies that the device needs to have a modulation bandwidth considerably exceeding the signal bandwidth in order to show significant squeezing. For lasers, a large bandwidth is normally achieved by operating the laser with a high rate of stimulated emission. Due to the small gain, however, nanolasers tend to be overdamped \cite{Moelbjerg2013,Romeira2018}. On the other hand, Purcell-enhanced nanoLEDs may feature a large modulation bandwidth \cite{Suhr2010} when exploiting extreme dielectric confinement \cite{Wang2018}. This is key to realizing devices that generate squeezed light within a bandwidth of several gigahertz. 

The pumping rate for the nanoLED corresponds to a current density less than 10 ${\rm kA/cm^2}$ using the small mode volume of Ref. \cite{Wang2018}, which is high, but feasible \cite{Khurgin2014}. Compared to plasmonic structures for achieving extreme confinement \cite{Khurgin2014}, there are no intrinsic losses for dielectric structures and for the high efficiency nanoLED structures considered here, the heat dissipated can be low or even negative due to the energy carried by the photons \cite{Li2019a}.

\begin{figure}[ht!]
	\includegraphics[width=0.28\textwidth]{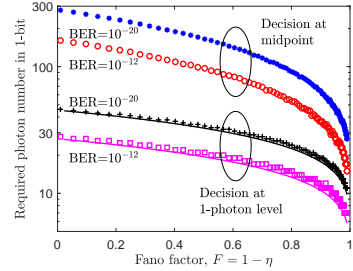}
		\caption{Required average number of photons in 1-bit to reach specified BER-values versus Fano factor. Upper curves are for a decision point at half the average number of photons, and lower curves are for a decision point at the 1-photon level. Solid curves are analytical estimates from Ref. \onlinecite{Saleh1992}}
	\label{fig:receiver_sensitivity}
\end{figure}

Fig. \ref{fig:receiver_sensitivity} shows the number of photons required in the 1-bit of an on-off keying system versus $F=1-\eta$ to achieve  bit-error-ratios (BERs) of $10^{-12}$ and $10^{-20}$ for a decision threshold at the 1-photon level (lower curves) and at the midpoint (upper curves). Notice that   communication links between the cores of a computer requires much lower BER than conventional communication systems. The curves show the very significant reduction in power level that can be obtained by reducing the Fano factor.

In conclusion, we have shown that nanoLEDs with near-unity efficiency are promising sources of intensity-squeezed light in a bandwidth of several gigahertz when pumped by a "quiet" current source and exploiting new designs for achieving extreme dielectric confinement \cite{Choi2017,Wang2018,Hu2018c}. On the other hand, nanolasers are fundamentally limited by the internal losses. 
We have also introduced an efficient approach for stochastic simulations of intenity-noise squeezing, which may be used for further theoretical investigations, e.g. addressing details of the pumping scheme.


 
\bibliography{bibliography_squeezing}

\end{document}